\documentclass{article}
\usepackage[]{epsfig}
\usepackage[centertags]{amsmath}
\usepackage{amsfonts}
\usepackage{amssymb}
\usepackage[english]{babel}
\usepackage{amsmath, amsthm, slashed,url}
\usepackage{color}
\usepackage{cite}
\usepackage{hyperref}

\begin{document}

%%%%%%%%%%%%%%%%%%%%%%%%%%%%%%%%%%%%%%%%%%%%%%%%%%%%%%%%%%%%%%%%%%%
\title{\bf Dynamic and static properties of Quantum Hall and Harmonic Oscillator systems on the non-commutative plane}

\author{ Nicol\'as Nessi and
Lucas~Sourrouille
\\
{\normalsize\it  IFLP-CONICET
}\\ {\normalsize\it Diagonal 113 y 64 (1900), La Plata, Buenos Aires, Argentina}
\\
{\footnotesize  lsourrouille@iflp.unlp.edu.ar} } \maketitle

\abstract{We study two quantum mechanical systems on the noncommutative plane using a representation independent approach. First, in the context of the Landau problem, we obtain an explicit expression for the gauge transformation that connects the Landau and the symmetric gauge in noncommutative space. This lead us to conclude that the usual form of the symmetric gauge 
$\vec{A}=\left(-\frac{\beta}{2}\hat{Y},\frac{\beta}{2}\hat{X}\right)$, in which the constant $\beta$ is interpreted as the magnetic field, is not true in noncommutative space. We also be able to establish a precise definition of $\beta$ as function of the magnetic field, for which the equivalence between the symmetric and Landau gauges is hold in noncommutative plane. 
Using the symmetric gauge we obtain results for the spectrum of the Quantum Hall system, its transverse conductivity in the presence of an electric field and other static observables. These results amend the literature on Quantum Hall Effect in noncommutative plane in which the incorrect form of the symmetric gauge, in noncommutative space, is assumed. We also study the non-equilibrium dynamics of simple observables for this system. On the other hand, we study the dynamics of the harmonic oscillator in non-commutative space and show that, in general, it exhibit quasi-periodic behavior, in striking contrast with its commutative version. The study of the dynamics reveals itself as a most powerful tool to characterize and understand the effects of non-commutativity.}

\vspace{0.3cm}
{\bf Keywords}: Quantum Mechanics; Noncommutative field theory; Landau Levels; Quantum Hall Effect

%%%%%%%%%%%%%%%%%%%%%%%%%%%%%%%%%%%%%%%%

%%%%%%%%%%%%%%%%%%%%%%%%%%%%%%%%%%%%%%%%
\vspace{1cm}
\section{Introduction}
\label{sec:intro}
The existence of a natural minimum length-scale which would function as a short-distance cutoff and eliminate the ultraviolet divergencies of quantum field theories was first advocated by Heisenberg in the 1930s~\cite{hossenfelder}. He was the first one to propose that such minimal length-scale could be implemented by means of non-commuting coordinates. The first concrete realization of non-commutative coordinates was first introduced by Snyder several years later~\cite{snyder}. This approach was abandoned with the advent of renormalisable field theories. However, these ideas regained relevance in the quest to formulate a quantum theory of gravity. In fact, the existence of a minimum length-scale below which position cannot be resolved is one of the characteristics that a quantum theory of gravity should incorporate~\cite{doplicher,calmet,hossenfelder,garay}.

%The Quantum Hall Effect (QHE) is one of the most remarkable condensed
%matter phenomena discovered in the second half of the 20 th century \cite{1}. The Hall Effect is a resistance in the transverse direction of the current in a 2D conductor that %emerges when a
%magnetic field is applied perpendicular to the conductor \cite{2}.
%The Integer Quantum Hall (IQH) Effect shows that this
%resistivity is quantized to extraordinary precision in integer multiples of fundamental flux when the conductor is
%brought to very low temperatures. A satisfactory model
%of IQH states emerges from a non-interacting model of
%the electrons in the conductor and a band structure of
%these states can be found \cite{3,4,5,6}. In this sences was  Laughlin the first to give an
%elegant and general explanation of the phenomenon, for a two-dimensional sample in a strong magnetic field \cite{7} see also \cite{8,9}.

In particular, the study of field theories in noncommutative space has received much attention in the last years in connection with string theory~\cite{douglas,szabo,schaposnik}. This connection was first considered by Connes et al.~\cite{connes}, who observed that noncommutative geometry arise as a possible scenario for certain low energy limits of string theory and M-theory. Afterwards, Seiberg and Witten showed that the low energy dynamics of string theory can be described in terms of the non-commutative Yang-Mills theory~\cite{seiberg}. Among the numerous application of non-commutative field theories that had been proposed since then, the description of the quantum Hall effect by means of the non-commutative Chern-Simons theory in 2+1 dimensions by Susskind is one of the most striking ones~\cite{susskind}.

There has been also studies on the effect of coordinate non-commutativity in other fields, like quantum electrodynamics~\cite{chaichian,chair}, cosmology~\cite{garcia,koh} and many-body theory~\cite{scholtz08}. The study of the non-commutative version of quantum mechanics has been also the object of an intense theoretical activity, see Refs.~\cite{gamboa,nair,li,scholtz,gangopadhyay,thom} among others. In this paper, we address some static properties of the Landau problem using a representation independent approach, reobtaining previous results and presenting new ones. Several articles concerned with the Landau problem or the closely related integer quantum Hall effect in non-commutative space were written. However many of them arrive to incorrect results both in the energy spectrum and the transverse conductivity due to an incorrect definition of the magnetic field and the symmetric gauge, see for example~\cite{dayi,basu,dulat,jiang,harms}. In this note we clarify these aspects providing the right definition of the magnetic field as well as the symmetric and Landau gauges so that they are gauge equivalent in noncommutative space.

On the other hand, the dynamics of even simple systems on the non-commutative plane has not been addressed yet. In this contribution, we study some aspects of the non-equilibrium dynamics of the Landau problem and the isotropic harmonic oscillator in non-commutative space. We find that the introduction of non-commutative coordinates does not affect qualitatively the dynamics of the Landau problem, while the picture is completely the opposite for the harmonic oscillator. For the latter, the presence of non-commutative coordinates introduces two radical modifications: (i) a coupling between the movements in the $x$ and $y$ directions, and (ii) the emergence of a splitting in the frequency of oscillation of the harmonic modes which causes the temporal evolution of the average of simple observables to exhibit quasiperiodicity. In order to present these results, we organised the paper as follows. In Sec.~\ref{sec:gauge} we show how to explicitly construct the gauge transformation linking the Landau and symmetric gauges in non-commutative spaces, and consequently obtain an expression for the latter. In Sec.~\ref{sec:landau_diagonalisation} we show a way to solve the Landau problem in the symmetric gauge in non-commutative space using a representation independent approach. In Sec.~\ref{sec:landau_dynamics} we use this solution to calculate the non-equilibrium dynamics of a charged particle in a magnetic field on the non-commutative plane. In Sec.~\ref{sec:qhe} we revisit the calculation of the Hall conductivity using a representation independent approach and the correct form of the symmetric gauge. In Sec.~\ref{sec:ho} we review some static results around the isotropic harmonic oscillator and calculate its non-equilibrium dynamics starting from a coherent initial state in order to illustrate some important differences with respect to the commutative case. Finally, in Sec.~\ref{sec:conclusions} we summarize the main results.

\section{A charged particle in a magnetic field}
\label{sec:landau}

\subsection{Gauge transformations in non-commutative space}
\label{sec:gauge}
The most general gauge transformation can be written as,
\begin{equation}
A^{\prime}_i=U A_i U^{\dagger}-\frac{\hbar}{i e}\,U\,\partial_i\,U^{\dagger},
\end{equation}
where
\begin{equation}
U=\exp\left(i\frac{e}{\hbar}\alpha\right),
\end{equation}
being $\alpha$ a function of the coordinates that defines the gauge transformation. For a commutative space, this form reduces to the usual gauge transformation, $A_i^{\prime}=A_i+\partial_i\,\alpha$. However, in non-abelian gauge theories or in our case in which we have non-commuting coordinates, the full form has to be considered.

We will go over the gauge transformation between the Landau and the symmetric gauge in the commutative space in order to set the stage for the non-commutative calculation. The Landau gauge is given by $\vec{A}=\left(0,B\,x\right)$,  where $B$ is the magnetic field. If we chose $\alpha=\frac{B}{2}x\,y$ we arrive to the symmetric gauge, $\vec{A}^{\prime}=\left(-\frac{B}{2}\,y,\frac{B}{2}\,x\right)$.

We will denote the non-commuting coordinates with capital letters and a hat, $\hat{X}$ and $\hat{Y}$,
\begin{equation}
\left[\hat{X},\hat{Y}\right]=i\theta.
\end{equation}
The Landau gauge assumes the same form, $\vec{A}=\left(0,B\,\hat{X}\right)$, and for the non-commutative gauge transformation we propose,
\begin{equation}
U=\exp\left(  i\,C\,\hat{X}\hat{Y}  \right),
\end{equation}
with $C$ a constant to be determined. In order to calculate derivatives we will use the commutators with the coordinates,
\begin{eqnarray}
\label{eq:nc_derivatives}% \nonumber % Remove numbering (before each equation)
  \partial_X\,U &=& -\frac{1}{i\theta}\left[\hat{Y},U\right], \\
  \partial_Y\,U &=& \frac{1}{i\theta}\left[\hat{X},U\right].
\end{eqnarray}
It is easy to show that,
\begin{eqnarray}
\label{eq:nc_transformation}
U\,\hat{X}\,U^{\dagger}&=&e^{iC\hat{X}\hat{Y}}\,\hat{X}\,e^{-iC\hat{X}\hat{Y}}=\hat{X}\,e^{C\theta},\\
U\,\hat{Y}\,U^{\dagger}&=&e^{iC\hat{X}\hat{Y}}\,\hat{Y}\,e^{-iC\hat{X}\hat{Y}}=\hat{Y}\,e^{-C\theta}.
\end{eqnarray}
Using Eq.~(\ref{eq:nc_derivatives}) and Eq.~(\ref{eq:nc_transformation}) we can derive the following expressions,
\begin{eqnarray}
U\,\partial_X\,U^{\dagger}&=&-\frac{1}{i\,\theta}\left(  e^{-C\theta}  -1\right)\hat{Y},\\
U\,\partial_Y\,U^{\dagger}&=&-\frac{1}{i\,\theta}\left(  e^{C\theta}  -1\right)\hat{X}.
\end{eqnarray}

Now we can calculate the transformed gauge field,
\begin{eqnarray}
A^{\prime}_X&=&-\frac{\hbar}{e\theta}\,\left(e^{-C\theta}-1\right)\,\hat{Y},\\
A^{\prime}_Y&=&\frac{\hbar}{e\theta}\,\left(\frac{e}{\hbar}\theta B\,e^{C\theta}+e^{C\theta}-1\right)\,\hat{X}.
\end{eqnarray}

The symmetric gauge is obtained by choosing $C$ in such a way that,
\begin{equation}
\frac{\hbar}{e\theta}\,\left(e^{-C\theta}-1\right)=\frac{\hbar}{e\theta}\,\left(\frac{e}{\hbar}\theta B\,e^{C\theta}+e^{C\theta}-1\right).
\end{equation}
The solution is,
\begin{equation}
e^{C\theta}=\frac{1}{\sqrt{1+\frac{e}{\hbar}\theta B}},
\end{equation}

From this equation we can deduce the value of the constant $C$ by applying natural logarithm 
to the both side of the equation. Them the constant $C$ is determined in terms on the parameter $\theta$ and the magnetic field $B$
and implies that the symmetric gauge takes the form,
\begin{equation}
\vec{A}^{\prime}=\left(-\frac{\beta}{2}\hat{Y},\frac{\beta}{2}\hat{X}\right),
\label{15}
\end{equation}
where
\begin{equation}
\beta=\frac{2\hbar}{e\theta}\left(-1+\sqrt{1+\frac{e}{\hbar}\theta\,B}\right).
\label{be}
\end{equation}

Now we will show explicitly that both gauges generate the same magnetic field. In order to define the magnetic field properly, we need to introduce momentum operators $\hat{P}_x$ and $\hat{P}_y$ satisfying usual commutation relations between themselves and the coordinates, such that we can construct the gauge invariant momentum operators,
\begin{equation}
\vec{\pi}=\vec{P}-e\,\vec{A},
\end{equation}
which is equivalent to the covariant derivative in quantum field theory. The magnetic field is then given by $B=\frac{1}{i\hbar e}\left[\pi_x,\pi_y\right]$. For the Landau gauge we obtain,
\begin{equation}
\left[\pi_x,\pi_y\right]=i\hbar e\,B,
\label{B}
\end{equation}
while for the symmetric gauge calculated above we get,
\begin{equation}
\left[\pi_x,\pi_y\right]=i\hbar e\,\beta\left(  1+\frac{e\beta}{4\hbar}\theta  \right)=i\hbar e\,B,
\label{b1}
\end{equation}
verifying explicitly that both gauges are equivalent. We also note that $\lim_{\theta\rightarrow 0}\beta=B$, so that the non-commutative symmetric gauge has the correct commutative limit.
The parameter $\beta$ does not be interpreted as a magnetic field $B$ unless $\theta \rightarrow 0$. When $\theta\rightarrow 0$, $\beta$ becomes equal to $B$ (this clare from equation (\ref{b1}. When this occur the Symmetric gauge (\ref{15}) coincides with the usual form of the symmetric gauge in the commutative space. It is intersting to remark from equation (\ref{be}) that $\beta$ becomes $\beta = \frac{-2\hbar}{e \theta}$ for $\theta = \frac{-\hbar}{e \theta}$. This is the minimum value of $\beta$ and stablish a restiction on $\theta$, since for $\theta < \frac{-\hbar}{e \theta}$, $\beta$ takes inmaginary values and therefore the gauge field \ref{15} would be imaginary. 
It is important to point out that in some part of the literature on the non-commutative Landau problem and non-commutative Hall effect, a naive version of the symmetric gauge is utilized \cite{dayi,basu,dulat,jiang,harms},
\begin{equation}
\vec{A}=\left(-\frac{B}{2}\hat{Y},\frac{B}{2}\hat{X}\right)
\label{20}
\end{equation}
 This leads us to a magnetic field of the form
\begin{equation}
\left[\pi_x,\pi_y\right]=i\hbar e\,B\left(  1+\frac{eB}{4\hbar}\theta  \right)
\end{equation}
which, clearly, is not the same as \ref{B}.
Therefore the gauge field \ref{20} is not equivalent to the Landau gauge, i.e., it does not describe the physics of a particle on a magnetic field of strength $B$. This fact led to the authors of references \cite{dayi,basu,dulat,jiang,harms} to wrong results regarding the energy spectrum and Hall conductivity. We will show in section \ref{sec:qhe} the correct calculation of this observables by using the gauge field \ref{15}.

\subsection{Diagonalisation of the Hamiltonian}
\label{sec:landau_diagonalisation}

We will address the Landau problem on the non-commutative plane. We need to consider a charged particle subject to an uniform perpendicular magnetic field. We can write the Hamiltonian of the system as,
\begin{equation}
\hat{H}=\frac{\pi_x^2}{2m}+\frac{\pi_y^2}{2m}%-e E \hat{X},
\end{equation}
using the results from the previous section, in the symmetric gauge the Hamiltonian takes the form,
\begin{equation}
\hat{H}=\frac{\left(  \hat{P}_x+e\,\beta\,\frac{\hat{Y}}{2}  \right)^2}{2m}+\frac{\left(  \hat{P}_y-e\,\beta\,\frac{\hat{X}}{2}  \right)^2}{2m},%-e E \hat{X}.
\end{equation}
The operators $\hat{X}$, $\hat{Y}$, $\hat{P}_x$ and $\hat{P}_y$ satisfy the algebra,
\begin{eqnarray}
\label{eq:algebra}
\nonumber \left[\hat{X},\hat{Y}\right]&=&i\theta,\\
\nonumber \left[\hat{P}_x,\hat{P}_y\right]&=&0,\\
\left[\hat{X},\hat{P}_x\right]&=&\left[\hat{Y},\hat{P}_y\right]=i\hbar,
\end{eqnarray}
where $\theta$ is the non-commutativity parameter.

%In order to diagonalize the Hamiltonian we start by defining the velocity operators,
%\begin{eqnarray}
%\hat{V}_x&=&\frac{\hat{P}_x+e\,B\,\frac{\hat{Y}}{2}}{m},\\
%\hat{V}_y&=&\frac{\hat{P}_y-e\,B\,\frac{\hat{X}}{2}}{m}.
%\end{eqnarray}
%The commutation relations between the velocity operators can be calculated from the algebra of operators in Eq.~(\ref{eq:algebra}),
%\begin{equation}
%[\hat{V}_x,\hat{V}_y]=i\frac{\hbar}{m}\omega_A,
%\end{equation}
%where,
%\begin{equation}
%\omega_A=\frac{e\,B}{m}\left(  1+\frac{e\,B}{4\,\hbar}\theta  \right).
%\end{equation}
In order to diagonalize the Hamiltonian, we can define creation and annihilation operators as,
\begin{eqnarray}
\hat{A}&=&\frac{\pi_x+i\,\pi_y}{\sqrt{2\hbar e B}},
\label{A1}
\end{eqnarray}
\begin{eqnarray}
\hat{A}^{\dag}&=&\frac{\pi_x-i\,\pi_y}{\sqrt{2\hbar B}},
\label{A2}
\end{eqnarray}
with the commutator given by,
\begin{equation}
\left[\hat{A},\hat{A}^{\dag}\right]=1.
\end{equation}

In terms of these operators the Hamiltonian acquires diagonal form,
\begin{equation}
\label{eq:ham_landau_diagonal}
\hat{H}=\hbar\,\omega_c\,\left(\hat{A}^{\dagger}\hat{A}+\frac{1}{2}\right)%-e E \hat{X},
\end{equation}
where $\omega_c=\frac{e\,B}{m}$ is the cyclotron frequency.

%It is clear that the spectrum of the Hamiltonian is of the harmonic oscillator type, $\hbar\,\omega_A\,\left(N+\frac{1}{2}\right)$, where $N$ is a non-negative integer. Also note that $\omega_A$ tends to the cyclotron frequency $\frac{e\,B}{m}$ when $\theta\rightarrow 0$, which provides the correct commutative limit.

We additionally introduce~\cite{jiang},
\begin{eqnarray}
\tilde{\pi}_x&=&\hat{P}_x-\sigma\,\hat{Y},\\
\tilde{\pi}_y&=&\hat{P}_y+\sigma\,\hat{X},
\end{eqnarray}
where
\begin{equation}
\sigma=\frac{e\,\beta/2}{1+\frac{e\,\beta}{2\,\hbar}\theta}=\frac{\hbar}{\theta}\left[1-\left(  1+\frac{e}{\hbar}\theta\,B  \right)^{-1/2}\right].
\end{equation}
The commutator of these operators is given by,
\begin{equation}
[\tilde{\pi}_x,\tilde{\pi}_y]=(-i\,\hbar)\left(  2\sigma-\sigma^2\frac{\theta}{\hbar}  \right)=(-i\hbar)\frac{e\,B}{1+\frac{e}{\hbar}\theta\,B}.
\end{equation}
We now define another set of creation and annihilation operators,
\begin{eqnarray}
\hat{B}&=&\sqrt{\frac{1+\frac{e}{\hbar}\theta\,B}{2e\hbar B}}\left(\tilde{\pi}_y+i\,\tilde{\pi}_x\right),\\
\hat{B}^{\dag}&=&\sqrt{\frac{1+\frac{e}{\hbar}\theta\,B}{2e\hbar B}}\left(\tilde{\pi}_y-i\,\tilde{\pi}_x\right),
\end{eqnarray}
which satisfy the following commutation relations,
\begin{eqnarray}
\left[\hat{B},\hat{B}^{\dag}\right]&=&1,\\
\left[\hat{A},\hat{B}^{\dag}\right]&=&\left[\hat{A},\hat{B}\right]=0.
\end{eqnarray}

The operator $\hat{B}^{\dagger}\hat{B}$ trivially commutes with the Hamiltonian, providing a new quantum number. Therefore, the eigenstates of the Hamiltonian are given by,
\begin{equation}
\vert n\,m\rangle=\vert n\rangle_{A}\otimes\vert m\rangle_{B}=\frac{\left(  \hat{A}^{\dagger}  \right)^n\,\left(  \hat{B}^{\dagger}  \right)^m}{\sqrt{n!\,m!}}\vert 0\rangle_{A}\otimes\vert 0\rangle_{B},
\end{equation}
where $\vert 0\rangle_{A}\otimes\vert 0\rangle_{B}$ is the state annihilated both by $\hat{A}$ and $\hat{B}$, the lowest Landau level. Just as in the commutative case, the eigenstates of the Hamiltonian are infinitely degenarate. The energy spectrum, composed of equidistant Landau levels, is the same as in the commutative space, $E_{n,m}=\hbar\omega_c (n+1/2)$~\cite{nair}. Notice that this results is at variance to what is found when the proper form of the symmetric gauge is not used~\cite{dayi,dulat,jiang}.

Now, we will show that the Hamiltonian eigenstates $\vert n\,m\rangle$ are also eigenstates of the angular momentum $L_z$. On the NC plane the angular momentum takes the form~\cite{scholtz},
\begin{equation}
\hat{L}_z=\hat{X}\hat{P}_y-\hat{Y}\hat{P}_x+\frac{\theta}{2\hbar}\left(  \hat{P}^2_{x}+\hat{P}^2_{y}  \right).
\end{equation}
Using the expression of $\hat{X}$, $\hat{Y}$, $\hat{P}_x$ and $\hat{P}_y$ in terms of the creation and annihilation operators,
\begin{eqnarray}
\label{eq:x_y_px_py_ab}
\nonumber \hat{X}&=&\sqrt{\frac{\hbar}{2eB}}\left[  \hat{B}+\hat{B}^{\dagger}+i\left(\hat{A}-\hat{A}^{\dagger}\right)\sqrt{1+\frac{e}{\hbar}\theta B}  \right],\\
\nonumber \hat{Y}&=&\sqrt{\frac{\hbar}{2eB}}\left[  i\left(\hat{B}-\hat{B}^{\dagger}\right)+\left(\hat{A}+\hat{A}^{\dagger}\right)\sqrt{1+\frac{e}{\hbar}\theta B}  \right],\\
\nonumber \hat{P}_x&=&\sqrt{\frac{\hbar}{2eB}}\frac{e\beta}{2}\left[ -i\left(\hat{B}-\hat{B}^{\dagger}\right)+\left(\hat{A}+\hat{A}^{\dagger}\right)  \right],\\
\hat{P}_y&=&\sqrt{\frac{\hbar}{2eB}}\frac{e\beta}{2}\left[ \left(\hat{B}+\hat{B}^{\dagger}\right)-i\left(\hat{A}-\hat{A}^{\dagger}\right)  \right],
\end{eqnarray}
we find an expression of $L_z$ in terms of $\hat{A},\hat{A}^{\dagger}$ and $\hat{B},\hat{B}^{\dagger}$,
\begin{equation}
\hat{L}_z=\hbar\left(  \hat{B}^{\dagger}\hat{B}-\hat{A}^{\dagger}\hat{A}  \right).
\end{equation}
This implies that $[H,\hat{L}_z]=0$, which is what we wanted to prove. In particular, $\hat{L}_z\,\vert 0\,0\rangle=0$.

At this point it is also possible to calculate the magnetic length $\ell^2=\langle \hat{X}^2\rangle+\langle \hat{Y}^2 \rangle$, which represents the average spatial spread of the state $\vert n\,m\rangle$. We obtain
\begin{equation}
\ell^2=\frac{2 \hbar}{e\,B}\left(  1+n+m  \right)+\theta(1+2n),
\end{equation}
where the first term is the commutative result. We see that space non-commutativity generates an additive extra contribution to the size of the wavefunction. In particular, $\theta$ sets the minimum spread of the wavefunction when $B\rightarrow\infty$.

%In order to obtain an explicit expression for the LLL wavefunction $\Psi^{(0,0)}$ we need to introduce a representation for the operators $\hat{X}$, $\hat{Y}$, $\hat{P}_x$ and $\hat{P}_y$. We shall follow the representation proposed by Scholtz et al. (cita). We give the main elements of the formalism in Appendix~\ref{app:representation}, in order to provide a self-contained presentation. For more details check Refs.~(citar scholtz y la tesis de thom).

\subsection{Dynamics}
\label{sec:landau_dynamics}
In this section we will solve the Heisenberg equations of motion of the system and analyse its dynamics. The dynamics of the $A$ and $B$ operators under Hamiltonian Eq.~(\ref{eq:ham_landau_diagonal}) is quite simple,
\begin{eqnarray}
\hat{A}(t)&=&\hat{A}\,e^{-i \omega_c t},\\
\hat{B}(t)&=&\hat{B}.
\end{eqnarray}
Using this expressions and the relations \ref{A1} and \ref{A2}, we can calculate the dynamics of $\hat{X}$ and $\hat{Y}$,
\begin{eqnarray}
\label{eq:dyn_landau}
\hat{X}(t)&=&\frac{\gamma}{e B}\left(\hat{P}_y+\sigma\hat{X}\right)+\left[  \hat{X}-\frac{\gamma}{eB}\left(  \hat{P}_y+\sigma \hat{X}  \right)  \right]\cos(\omega_c t)
+\left[  \hat{Y}+\frac{\gamma}{eB}\left(  \hat{P}_x-\sigma \hat{Y}  \right)  \right]\sin(\omega_c t),\\
\nonumber \hat{Y}(t)&=&-\frac{\gamma}{e B}\left(\hat{P}_x-\sigma\hat{Y}\right)+\left[  \hat{Y}+\frac{\gamma}{eB}\left(  \hat{P}_x-\sigma \hat{Y}  \right)  \right]\cos(\omega_c t)
+\left[  -\hat{X}+\frac{\gamma}{eB}\left(  \hat{P}_y+\sigma \hat{X}  \right)  \right]\sin(\omega_c t),
\end{eqnarray}
where we defined,
\begin{equation}
\gamma=\sqrt{1+eB\frac{\theta}{\hbar}}.
\end{equation}

As expected, the magnetic field introduces a coupling between the movements in the $x$ and $y$ directions. Notice that in the limit of $\theta\rightarrow 0$ we recover the commutative space results by the replacements $\gamma\rightarrow 1$ and $\sigma\rightarrow \frac{eB}{2}$. The non-commutativity of the coordinates does not have a decisive effect in this case, as a non-zero value of $\theta$ only renormalises the coefficients but does not change the structure of the solutions.

To illustrate this we will visualise the dynamics starting from a superposition state,
\begin{equation}
\vert\Psi_0\rangle=\frac{1}{2}\,\left(\vert 0\rangle+\vert 1 \rangle\right)_A\otimes\left(\vert 0\rangle+\vert 1 \rangle\right)_B.
\end{equation}
With this initial state the evolution of the averages takes the simple form,
\begin{eqnarray}
\langle \hat{X}(t)\rangle&=&\sqrt{\frac{\hbar}{2eB}}+\sqrt{\frac{\hbar}{2eB}}\gamma\,\sin(\omega_c t),\\
\langle \hat{Y}(t)\rangle&=&\sqrt{\frac{\hbar}{2eB}}\gamma\,\cos(\omega_c t),
\end{eqnarray}
where $\langle\cdots\rangle$ denotes average with respect to $\vert\Psi_0\rangle$. The averages describe a circle of square radius $\frac{\hbar}{2Be}\gamma^2=\frac{\hbar}{2Be}+\frac{\theta}{2}$, which has a minimum value given by $\theta/2$ when $B\rightarrow\infty
$, in the same fashion as the magnetic length.

In conclusion, the non-commutativity does not affect the overall behavior of the dynamics but only introduces quantitative changes in the parameters. In the next section we will show that for the harmonic oscillator the picture is completely different.

\subsection{Transverse stationary currents}
\label{sec:qhe}
One of the most striking manifestations of the coupling between the $x$ and $y$ directions in the presence of a magnetic field is the appearance of a non-vanishing current when applying an electric field in one of the directions. In fact, the transverse conductivity is the main observable in the quantum Hall effect experiments. In this section we will study the effect of the non-commutativity of the coordinates on this current. This calculation has been undertaken several times in the past, but in many cases the results are not correct due to the utilization of the wrong gauge as we explained earlier. Moreover, our approach does not depend on a particular representation of the operator algebra.

The Hamiltonian is,
\begin{equation}
\hat{H}=\frac{\left(  \hat{P}_x+e\,\beta\,\frac{\hat{Y}}{2}  \right)^2}{2m}+\frac{\left(  \hat{P}_y-e\,\beta\,\frac{\hat{X}}{2}  \right)^2}{2m}-e E \hat{X}.
\end{equation}

Using the expression of $\hat{X}$ in terms of the creation and annihilation operators, the Hamiltonian may be expressed as,
\begin{equation}
\label{eq:hamiltonian_diagonal}
H=\hbar\omega_c\left[  \left(  \hat{A}^{\dagger}+\lambda^{*}  \right)\left(  \hat{A}+\lambda  \right) - \vert\lambda\vert^2 + \frac{1}{2}  \right]-e E \sqrt{\frac{\hbar}{2eB}}\left(\hat{B}+\hat{B}^{\dagger}\right),
\end{equation}
where
\begin{equation}
\lambda=i\frac{e E}{\hbar \omega_c}\sqrt{\frac{\hbar}{2 e B}}\sqrt{1+\frac{e}{\hbar}\theta B}.
\label{45}
\end{equation}

%This implies that the operator $\hat{B}^{\dagger}\hat{B}$ commutes with the magnetic part of the Hamiltonian, providing a new quantum number. Therefore, the eigenstates of the Hamiltonian are given by,
%\begin{equation}
%\Psi^{(n,m)}=\frac{\left(  \hat{A}^{\dagger}  \right)^n\,\left(  \hat{B}^{\dagger}  \right)^m}{\sqrt{n!\,m!}}\Psi^{(0,0)},
%\end{equation}
%where $\Psi^{(0,0)}$ is the state annihilated both by $\hat{A}$ and $\hat{B}$. Just as in the commutative case, the eigenstates of the Hamiltonian are infinitely degenarate.

The Hamiltonian Eq.~(\ref{eq:hamiltonian_diagonal}) can be divided into two commuting terms, $H_1=\hbar\omega_c\left[  \left(  \hat{A}^{\dagger}+\lambda^{*}  \right)\left(  \hat{A}+\lambda  \right) - \vert\lambda\vert^2 + \frac{1}{2}  \right]$ and $H_2=-e E \sqrt{\frac{\hbar}{2eB}}\left(\hat{B}+\hat{B}^{\dagger}\right)$. These two parts can be diagonalised separately and the eigenstates can be constructed as direct products of the eigenstates of the individual terms. Regarding $H_1$, we can construct its eigenstates using the displacement operator $D=\exp\left(  \lambda \hat{A}^{\dagger}-\lambda^{*}\hat{A}  \right)$ since,
\begin{equation}
\left(  \hat{A}^{\dagger}+\lambda^{*}  \right)\left(  \hat{A}+\lambda  \right)D^{\dagger}\vert n \rangle_{A}=n\,D^{\dagger}\vert n \rangle_{A}.
\end{equation}
To show this, it is enough to note that $D^{\dagger}\hat{A}^{\dagger}D=\hat{A}^{\dagger}+\lambda^{*}$ and $D^{\dagger}\hat{A}D=\hat{A}+\lambda$. Regarding $H_2$ we will limit ourselves to say that the eigenvalues are real since $\hat{B}+\hat{B}^{\dagger}$ is an hermitian operator,
\begin{equation}
\left(\hat{B}+\hat{B}^{\dagger}\right)\vert \xi \rangle_{B}=\xi \vert \xi \rangle_{B},\;\xi\in \mathbb{R}.
\end{equation}
Finally, the spectrum of the Hamiltonian is,
\begin{equation}
\label{eq:spectrum_nc_hall}
E_{n,\xi}=\hbar\,\omega_c\left(  k-\vert\lambda\vert^2+\frac{1}{2}  \right)-e\,E\sqrt{\frac{\hbar}{2eB}}\xi,
\end{equation}
whereas the eigenstates are $D^{\dagger}\vert n \rangle_{A} \bigotimes \vert \xi\rangle_{B}$. The only difference between the spectrum in Eq.~(\ref{eq:spectrum_nc_hall}) and the commutative spectrum, $\theta\rightarrow 0$, comes from the expression for $\lambda$ defined in formula Eq.~(\ref{45}).% In addition it is interesting to note that the landau levels (i.e. the spectrum \ref{eq:spectrum_nc_hall} for $E=0$) has no changes with respect to commutative case. This is a very significant diference with the energy spectrum found in the references \cite{dayi,basu,dulat,jiang,harms}, since the fact that the Landau Levels change due to noncommutativity of  space is one of the relevant result which these articles.

%Notice that the only difference between the spectrum in Eq.~(\ref{eq:spectrum_nc_hall}) and its commutative counterpart ($\theta\rightarrow 0$) is contained in the expression for $\lambda$.

We will calculate the average of the currents $\langle J_x\rangle$ and $\langle J_y\rangle$ with respect to arbitrary eigenstates of the Hamiltonian. The definition of the currents are,
\begin{eqnarray}
J_x&=&e\rho\,\frac{d\hat{X}}{dt}=\frac{e\rho}{i\hbar}\left[\hat{X},H\right],\\
J_y&=&e\rho\,\frac{d\hat{Y}}{dt}=\frac{e\rho}{i\hbar}\left[\hat{Y},H\right],
\end{eqnarray}
where $\rho$ is the electron density. We obtain for $J_x$,
\begin{equation}
J_x=e\rho\omega_c\sqrt{\frac{\hbar}{2eB}}\sqrt{1+\frac{e}{\hbar}\theta B}\left(  \hat{A}+\hat{A}^{\dagger}  \right).
\end{equation}

On the other hand, for $J_y$ we obtain,
\begin{equation}
J_y=\frac{e\rho}{i\hbar}\left[  \hbar\omega_c\sqrt{\frac{\hbar}{2eB}}\sqrt{1+\frac{e}{\hbar}\theta B}\left(  \hat{A}-\hat{A}^{\dagger}+2\lambda \right)  \right]-e\rho\frac{E}{B}.
\end{equation}

In order to evaluate the averages, we use that $\langle \hat{A}^{\dagger}\rangle=\lambda$ and $\langle \hat{A}\rangle=-\lambda$ to obtain,
\begin{eqnarray}
\langle J_x\rangle&=&0,\\
\langle J_y\rangle&=&-e\rho\frac{E}{B}.
\end{eqnarray}
The modifications introduced by the non-vanishing $\theta$ cancel out and the transverse conductivity $\sigma_{y}=\langle J_y\rangle/E$ is the same as in the commutative case. We stress that if the incorrect form of the gauge field is used, spurious modifications may appear in the transverse conductivity when $\theta \neq 0$.

\section{The isotropic harmonic oscillator}
\label{sec:ho}
In this section we will give a clear example in which the inclusion of non-commuting coordinates radically change the static and dynamic behavior of a simple system, namely the isotropic harmonic oscillator.

The Hamiltonian is,
\begin{equation}
\label{eq:ham_ho}
H=\frac{\hat{P}^2_x}{2m}+\frac{\hat{P}^2_y}{2m}+\frac{m\,\omega^2}{2}\left( \hat{X}^2+\hat{Y}^2\right),
\end{equation}
and the commutation relations are those in Eq.~(\ref{eq:algebra}). This Hamiltonian can be diagonalised using the following operators~\cite{scholtz}:
\begin{eqnarray}
\label{eq:ab}\hat{A}&=&\frac{1}{\sqrt{K_A}}\left(-\frac{\lambda_A}{\hbar}\hat{X}-i\frac{\lambda_A}{\hbar}-i \hat{P}_x+\hat{P}_y\right),\\
\nonumber\hat{B}&=&\frac{1}{\sqrt{K_B}}\left(\frac{\lambda_B}{\hbar}\hat{X}-i\frac{\lambda_B}{\hbar}+i \hat{P}_x+\hat{P}_y\right),
\end{eqnarray}
where the various constants are defined as:
\begin{eqnarray}
\lambda_{A}&=&\frac{1}{2}\left(  m^2\omega^2\,\theta + m\omega\,\sqrt{4\hbar^2+m^2\omega^2\theta^2}  \right),\\
\lambda_{B}&=&\frac{1}{2}\left(  -m^2\omega^2\,\theta + m\omega\,\sqrt{4\hbar^2+m^2\omega^2\theta^2}  \right),\\
K_{A}&=&\lambda_A\left(\frac{2\lambda_{A}\theta}{\hbar^2}+4\right),\\
K_{B}&=&\lambda_B\left(-\frac{2\lambda_{B}\theta}{\hbar^2}+4\right).
\end{eqnarray}

With these definitions it can be verified that the $\hat{A}$ and $\hat{B}$ operators satisfy a Fock algebra:
\begin{eqnarray}
\left[\hat{A},\hat{A}^{\dagger}\right]&=&\left[\hat{B},\hat{B}^{\dagger}\right]=1,\\
\left[\hat{A},\hat{B}^{\dagger}\right]&=&\left[\hat{A},\hat{B}\right]=0,
\end{eqnarray}
and the Hamiltonian acquires diagonal form:
\begin{equation}
H=\frac{\lambda_{A}}{m}\left( \hat{A}^{\dagger}\hat{A}+\frac{1}{2}\right)+\frac{\lambda_{B}}{m}\left( \hat{B}^{\dagger}\hat{B}+\frac{1}{2}\right).
\end{equation}
Therefore the spectrum is,
\begin{equation}
E_{n_A,m_B}=\frac{\lambda_{A}}{m}\left( n_A+\frac{1}{2}\right)+\frac{\lambda_{B}}{m}\left( m_B+\frac{1}{2}\right),
\end{equation}
with $n_A$ and $m_B$ non-negative integers. As was already known, the spectrum of the harmonic oscillator is modified for $\theta\neq 0$~\cite{scholtz,nair}. In particular, two different frequencies appear, which will have a profound impact on the dynamics, as we will see below.

To solve the Heisenberg equations of motion, we first notice that the dynamics of the Fock operators is a simple oscillatory factor,
\begin{eqnarray}
\hat{A}(t)&=&e^{-\frac{i}{\hbar}\frac{\lambda_A}{m}t}\hat{A},\\
\hat{B}(t)&=&e^{-\frac{i}{\hbar}\frac{\lambda_B}{m}t}\hat{B}.
\end{eqnarray}
Using this and the definitions in Eq.~(\ref{eq:ab}) we obtain the dynamics of $\hat{X}$ and $\hat{Y}$:
\begin{eqnarray}
\nonumber\hat{X}(t)&=&\frac{\hbar}{\lambda_A+\lambda_B}\bigg[\left(  \hat{X}\frac{\lambda_A}{\hbar}-\hat{P}_y  \right)\cos\left(\frac{\lambda_A}{\hbar m}t\right)+\left(  \hat{X}\frac{\lambda_B}{\hbar}+\hat{P}_y  \right)\cos\left(\frac{\lambda_B}{\hbar m}t\right)\\
\label{eq:xt}&+& \left(  \hat{P}_x+\hat{Y}\frac{\lambda_A}{\hbar}  \right)\sin\left(\frac{\lambda_A}{\hbar m}t\right)+\left(  \hat{P}_x-\hat{Y}\frac{\lambda_B}{\hbar}  \right)\sin\left(\frac{\lambda_B}{\hbar m}t\right)\bigg],
\end{eqnarray}
and,
\begin{eqnarray}
\nonumber\hat{Y}(t)&=&\frac{\hbar}{\lambda_A+\lambda_B}\bigg[\left(  \hat{Y}\frac{\lambda_A}{\hbar}+\hat{P}_x  \right)\cos\left(\frac{\lambda_A}{\hbar m}t\right)+\left(  \hat{Y}\frac{\lambda_B}{\hbar}-\hat{P}_x  \right)\cos\left(\frac{\lambda_B}{\hbar m}t\right)\\
\label{eq:yt}&+& \left(  \hat{P}_y-\hat{X}\frac{\lambda_A}{\hbar}  \right)\sin\left(\frac{\lambda_A}{\hbar m}t\right)+\left(  \hat{P}_y-\hat{X}\frac{\lambda_B}{\hbar}  \right)\sin\left(\frac{\lambda_B}{\hbar m}t\right)\bigg].
\end{eqnarray}
In the limit of $\theta\rightarrow 0$ we recover the known dependencies,
\begin{eqnarray}
\nonumber\hat{X}(t)&=&\hat{X}\cos(\omega t)+\frac{\hat{P}_x}{m\omega}\sin(\omega t),\\
\label{eq:xtyx_com}\hat{Y}(t)&=&\hat{Y}\cos(\omega t)+\frac{\hat{P}_y}{m\omega}\sin(\omega t).
\end{eqnarray}
Then, the first thing to notice from Eqs.~(\ref{eq:xt}) and~(\ref{eq:yt}) is that the movement in the $x$ and $y$ directions are coupled, as opposed to the commutative case, Eq.~(\ref{eq:xtyx_com}). The effect of non-commutativity changes dramatically the structure of the solutions. To illustrate this, we would like to construct an initial state for which $\langle\hat{Y}\rangle=\langle\hat{P}_y\rangle=0$. For such initial state, in commutative space, due to the fact that the movements in the $x$ and $y$ directions are decoupled, the evolution of the average $\langle \hat{Y}\rangle$ would be trivial, $\langle \hat{Y}\rangle=0$. However, in non-commutative space, both directions are coupled and movement in the $x$ direction induces movement in the $y$ direction, as we will see. To engineer such a state, we will use coherent states, $\vert\Psi_0\rangle=\vert\alpha\,\beta\rangle=\vert\alpha\rangle_{A}\otimes\vert\beta\rangle_{B}$ with,
\begin{eqnarray}
\vert\alpha\rangle_{A}&=&e^{-\vert\alpha\vert^2/2}\,e^{\hat{A}^{\dagger}}\vert0\rangle_{A},\\
\vert\beta\rangle_{B}&=&e^{-\vert\beta\vert^2/2}\,e^{\hat{B}^{\dagger}}\vert0\rangle_{B},
\end{eqnarray}
where $\alpha$ and $\beta$ are complex numbers and $\vert0\rangle_{A}$ and $\vert0\rangle_{B}$ is the state annihilated by $\hat{A}$ and $\hat{B}$, respectively. These states are eigenstates of the corresponding annihilation operators,
\begin{equation}
\hat{A}\vert\alpha\rangle_A=\alpha\vert\alpha\rangle_A.
\end{equation}
and analogously for $\hat{B}$. If we chose a particular coherent state as initial condition,
\begin{eqnarray}
\alpha&=&-\frac{\sqrt{K_B}\lambda_A}{\sqrt{K_A}\lambda_B},\\
\beta&=&1,
\end{eqnarray}
we have $\langle\hat{Y}\rangle=\langle\hat{P}_y\rangle=0$ as we wished, and $\langle\hat{P}_x\rangle=0$, $\langle\hat{X}\rangle=\frac{\hbar\sqrt{K_{B}}}{\lambda_A+\lambda_B}\left(   1+\frac{\lambda_A}{\lambda_B} \right)$.

With this initial condition, the evolution of the averages of $\hat{X}(t)$ and $\hat{Y}(t)$ is given by,
\begin{eqnarray}
\label{eq:dyn_coherent}
\langle\hat{X}(t)\rangle&=&\frac{\langle\hat{X}\rangle}{\lambda_A+\lambda_B}\left[  \lambda_A\cos\left(\frac{\lambda_A}{\hbar\,m}\,t\right)+\lambda_B\cos\left(\frac{\lambda_B}{\hbar\,m}\,t\right)  \right],\\
\langle\hat{Y}(t)\rangle&=&\frac{\langle\hat{X}\rangle}{\lambda_A+\lambda_B}\left[  \lambda_B\sin\left(\frac{\lambda_B}{\hbar\,m}\,t\right)-\lambda_A\cos\left(\frac{\lambda_A}{\hbar\,m}\,t\right)  \right].
\end{eqnarray}
We verify that the dynamics of $\langle\hat{Y}(t)\rangle$ for $\theta\neq 0$ is not trivial as expected. More interestingly, the presence of two different frequencies, $\frac{\lambda_A}{\hbar\,m}$ and $\frac{\lambda_B}{\hbar\,m}$, produces a series of crossovers in the dynamics as $\theta$ is varied. The basic observation is that if $\lambda_A/\lambda_B$ is a rational number the sum of harmonic functions with frequencies $\lambda_A$ and $\lambda_B$, respectively, will be a periodic function. On the contrary, if $\lambda_A/\lambda_B$ is an irrational number, the ensuing interference of harmonic functions with incommensurate frequencies gives rise to quasiperiodic behavior. In order to perform a quantitative analysis, we will restrict ourselves to small values of the non-commutativity parameter, $\theta\ll\frac{\hbar}{m\omega}$. In such regime,
\begin{equation}
\frac{\lambda_A}{\lambda_B}=1+\frac{m\omega}{\hbar\theta}+\mathcal{O}(\theta^2).
\end{equation}
Whenever $1+\frac{m\omega}{\hbar\theta}$ is close to a rational number, both frequencies $\lambda_A$ and $\lambda_B$ are commensurate and the temporal evolution of $\langle\hat{X}(t)\rangle$ and $\langle\hat{Y}(t)\rangle$ will be periodic. In particular, for small frequencies, $\lambda_A/\lambda_B\simeq 1$ and we will have the typical modulated oscillation pattern with an envelope of frequency $\vert \lambda_A-\lambda_B  \vert$, analogous to the acoustic beats. On the other hand, if $\frac{m\omega}{\hbar\theta}$ is an irrational number the temporal evolution of the averages will display a quasiperiodic pattern. Remarkably, we conclude that, in general, the dynamics of the harmonic oscillator in non-commutative space is not periodic.

In summary, the dynamics of the harmonic oscillator in non-commutative space is radically different from the commutative version, even for small $\theta$, due to the presence of a coupling between the dynamics in the $x$ and $y$ directions and the presence of two different frequencies $\lambda_A$ and $\lambda_B$. Even if these results were obtained for a very particular initial condition, the basic picture holds for more general initial states.

\section{Summary}
\label{sec:conclusions}
We have studied the non-commutative version of two of the simplest two-dimensional quantum systems, namely, the (Landau) problem of a particle in two dimensions subject to a transverse magnetic field and the two-dimensional harmonic oscillator. In the context of the Landau problem, we worked out explicitly the form of the gauge transformation the links the Landau and symmetric gauges. We obtained results for the spectrum and the transverse conductivity of the system in the presence of an electric field. Non-commutativity has no effect in these observables, a result that is at variance with what is obtained in some part of the literature in which an incorrect form of the symmetric gauge is utilized. We also showed explicitly that the eigenvector of the Hamiltonian in the symmetric gauge are also eigenstates of the angular momentum and calculated the corrections to the magnetic length due to non-commutativity. On the other hand, we showed that the dynamics of the system is not dramatically changed by non-commutativity, a non-vanishing $\theta$ only renormalizes the parameters that characterize the dynamics. The situation is completely the opposite for the harmonic oscillator. The dynamics of this system is profoundly affected by the fact that the coordinates do not commute. In particular we find that the movements in the $x$ and $y$ directions are coupled, something that we expected. More surprisingly, we find that the effect of a non-vanishing $\theta$ introduces an splitting in the harmonic oscillator frequency. The system has now two different natural frequencies even if the oscillator is isotropic. The interplay between both frequencies induces quasiperiodic behavior when they are incommensurate. This remarkable behavior shows the profound and unexpected influence that non-commutativity can have in quantum systems.

\vspace{0.6cm}

{\bf Acknowledgements}
\\
This work was supported by CONICET. We would like to express our gratitude to P. Pisani for discussions around the gauge transformation in non-commutative space.

\end{document}